\documentclass[reprint,showpacs,aps,prx,amsmath,amssymb,floatfix,superscriptaddress]{revtex4-2}
\usepackage{natbib}

\usepackage{graphicx}
\usepackage{siunitx}
\usepackage{physics}
\usepackage{MnSymbol}
\usepackage{mathtools}

\usepackage{hyperref}
\hypersetup{
    colorlinks=true,
    citecolor=blue,
    linkcolor=blue,
    anchorcolor=black,
    urlcolor=blue,
}
\usepackage{dcolumn} 
\usepackage{csquotes}
\usepackage{bm}
\usepackage{xcolor}
\usepackage{xspace} 
\usepackage{float}

\newcommand{\up}{\ensuremath{\ket \uparrow}\xspace}
\newcommand{\down}{\ensuremath{\ket \downarrow}\xspace}
\newcommand{\g}{\ensuremath{\ket g}\xspace}

\begin{document}
\title{Entanglement of mechanical oscillators mediated by a Rydberg tweezer chain}
\author{Cedric Wind}
\thanks{These authors contributed equally.}
\affiliation{Institut f\"{u}r Angewandte Physik, University of Bonn, Wegelerstr. 8, 53115 Bonn, Germany}
\author{Chris Nill}
\thanks{These authors contributed equally.}
\affiliation{Institut f\"{u}r Angewandte Physik, University of Bonn, Wegelerstr. 8, 53115 Bonn, Germany}
\affiliation{Institut f\"{u}r Theoretische Physik and Center for Integrated Quantum Science and Technology, Universität T\"{u}bingen, Auf der Morgenstelle 14, 72076 T\"{u}bingen, Germany}
\author{Julia Gamper}
\affiliation{Institut f\"{u}r Angewandte Physik, University of Bonn, Wegelerstr. 8, 53115 Bonn, Germany}
\author{Samuel Germer}
\affiliation{Institut f\"{u}r Angewandte Physik, University of Bonn, Wegelerstr. 8, 53115 Bonn, Germany}
\author{Valerie Mauth}
\affiliation{Institut f\"{u}r Angewandte Physik, University of Bonn, Wegelerstr. 8, 53115 Bonn, Germany}
\author{Wolfgang Alt}
\affiliation{Institut f\"{u}r Angewandte Physik, University of Bonn, Wegelerstr. 8, 53115 Bonn, Germany}
\author{Igor Lesanovsky}
\affiliation{Institut f\"{u}r Theoretische Physik and Center for Integrated Quantum Science and Technology, Universität T\"{u}bingen, Auf der Morgenstelle 14, 72076 T\"{u}bingen, Germany}
\affiliation{School of Physics and Astronomy and Centre for the Mathematics and Theoretical Physics of Quantum Non-Equilibrium Systems, The University of Nottingham, Nottingham, NG7 2RD, United Kingdom}
\author{Sebastian Hofferberth}
\affiliation{Institut f\"{u}r Angewandte Physik, University of Bonn, Wegelerstr. 8, 53115 Bonn, Germany}

\date{\today}

\begin{abstract}
Mechanical systems provide a unique test bed for studying quantum phenomena at macroscopic length scales.
However, realizing quantum states that feature quantum correlations among macroscopic mechanical objects remains an experimental challenge.
Here, we propose a quantum system in which two micro-electromechanical oscillators interact through a chain of Rydberg atoms confined in optical tweezers.
We demonstrate that the coherent dynamics of the system generate entanglement between the oscillators.
Furthermore, we utilize the tunability of the radiative decay of the Rydberg atoms for dissipative entanglement generation.
Our results highlight the potential to exploit the flexibility and tunability of Rydberg atom chains to generate nonclassical correlations between distant mechanical oscillators.
\end{abstract}

\maketitle

\section{Introduction}
Many experimental platforms explore entanglement and its applications using microscopic objects, such as photons~\cite{Aspect1982,Wang2016,Kam2025}, atoms~\cite{Vanleent2022,Saha2025}, and ions~\cite{Jost2009,Krutyanskiy2023}.
Extending entanglement to macroscopic objects, such as massive mechanical oscillators, is challenging, as decoherence usually becomes more pronounced with increasing system size~\cite{Zurek2003,Schlosshauer2019,Aspelmeyer2014,Liu2004}.
However, entangled macroscopic systems, and in particular mechanical oscillators, offer new opportunities for probing the quantum–classical boundary~\cite{Belli2016,Bild2023} and for exploring novel regimes of quantum acoustics~\cite{Renninger2018,Chu2020,Chegnizadeh2024,Rahman2025}. Moreover, they may serve as a test bed for models of quantum gravity~\cite{Bose2025, Anastopoulos2022, Bose2023}. 

So far, mechanical oscillators have predominantly been entangled using opto- and electromechanical coupling to cavities~\cite{Ockeloen-Korppi2018,Riedinger2018,MercierdeLepinay2021}.
A notable exception is a recent experiment, where micro-electromechanical oscillators with \si{\giga\hertz} resonance frequencies are coupled to superconducting qubits that mediate entanglement~\cite{Chou2025}. 
In a parallel development, micro-electromechanical oscillators have been realized that feature similar resonance frequencies, but offer dramatically enhanced coherence times on the order of one millisecond~\cite{Yang2024,Gokhale2020}. 

In this work, we propose to entangle two \si{\giga\hertz} micro‑electromechanical oscillators by coupling them to opposite ends of a chain of tweezer-trapped Rydberg atoms, as sketched in Fig. \ref{fig:cartoon}(a).
Rydberg atoms feature strong electric dipole transitions in the \si{\giga\hertz}-frequency regime that enable resonant interaction with the electromechanical oscillators. 
Moreover, trapping atoms in arrays of optical tweezer traps \cite{deLeseleuc2019,Bluvstein2023} allows for precise control of the interatomic distance as well as the distance between the atoms and the oscillators. 
Here, the spatial dependence of the underlying dipolar interactions permits control over the atom-atom and atom-oscillator coupling.
We build our investigation on a model that captures the essential coherent and dissipative processes of this hybrid system.
This model allows us to show that coherent excitation transport through the Rydberg chain mediates entanglement between the oscillators. 
Furthermore, we present a scheme for dissipative entanglement generation that exploits atomic decay processes.
The entanglement among the oscillators at the end of the dissipative dynamics is probabilistic. 
We discuss how the likelihood for obtaining entangled states can be enhanced through particular choices of the system parameters.

\section{Mechanical oscillators coupled by a Rydberg atom chain}
\begin{figure*}[t]
    \centering
    \includegraphics[width=\linewidth]{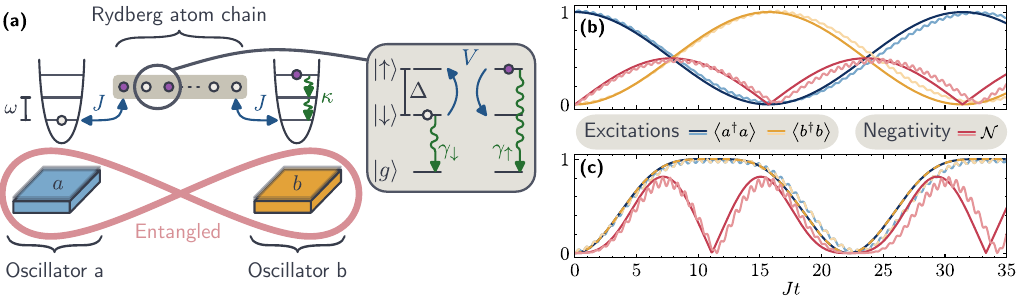}
    \caption{\textbf{Mechanical oscillators linked by a Rydberg atom chain.}
    \textbf{(a):~}
    Schematic of the oscillator-atom system illustrating the levels, coupling strengths and decay rates introduced in the main text. 
    Excitations in the Rydberg spin chain and the oscillators are depicted as violet filled circles, while white filled circles indicate the absence of excitations.
    \textbf{(b,c):~}Coherent evolution of the number of excitations in the oscillators $\ev{a^\dagger a}$, $\ev{b^\dagger b}$ and oscillator entanglement quantified by the negativity $\mathcal{N}$. The results correspond to the system presented in \textbf{(a)} with $N=2$ atoms and $V = 10J$.
    In panel \textbf{(b)} the system is initialized in $\ket{\psi_1}$ and in panel \textbf{(c)} in $\ket{\psi_2}$ (see text).
    Dark lines are obtained from the effective Hamiltonian~\eqref{eq:eff-hamiltonian}, while lighter, semi-transparent lines show the exact dynamics under the full Hamiltonian~\eqref{eq:total-hamiltonian}.
    In panel \textbf{(c)}, $\ev{b^\dagger b}$ is shown as a dashed curve to improve visibility.
    }
    \label{fig:cartoon}
\end{figure*}
\label{sec:system}
As depicted in Fig.~\ref{fig:cartoon}(a), the system we consider is composed of two micro-electromechanical oscillators, which are linked to the two ends of a chain of $N$ Rydberg atoms. 
The Hamiltonian of this system is given by 
\begin{equation}
    H=H_\mathrm{osc}+H_\mathrm{chain}+H_\mathrm{couple}.
    \label{eq:total-hamiltonian}
\end{equation}
Here, $H_\mathrm{osc}= \omega (a^\dagger a+b^\dagger b)$ is the Hamiltonian describing the relevant modes of the two oscillators with resonance frequency $\omega$ and corresponding ladder operators $a,a^\dagger$ and $b,b^\dagger$ respectively.
Throughout this work, we set $\hbar=1$.
We model the decoherence of the oscillator modes as an excitation decay with rate $\kappa$.
The system is considered to be in a zero-temperature environment, so that any heating from outside can be neglected (see Appendix~\ref{sec:zero-temperature}).

In the chain, each atom is modeled with two Rydberg states, denoted by the spin states \up and \down with energy separation $\Delta$.
We consider the two Rydberg states to spontaneously decay with rates $\gamma_\uparrow$ and $\gamma_\downarrow$ to an effective ground state $\ket g$, as illustrated in the inset in Fig.~\ref{fig:cartoon}(a).
This simplifies the decay dynamics of real Rydberg atoms but captures the essential aspect of the decay, namely the depopulation of the Rydberg state.
Due to the zero-temperature environment, blackbody-induced transitions to neighboring Rydberg states are not present.
We treat the interaction between the Rydberg atoms in the chain as dipolar flip-flop interaction between nearest neighbors with strength $V$~\cite{Barredo2015,deLeseleuc2019}, described by the Hamiltonian
\begin{equation}
    H_\mathrm{chain}=V\sum_{i=1}^{N-1} \left(\dyad{\downarrow}{\uparrow}_i\otimes \dyad{\uparrow}{\downarrow}_{i+1}+\mathrm{h.c.}\right)
    +\Delta \sum_{i=1}^N \mathcal P_\uparrow^{(i)},
    \label{eq:chain-hamiltonian}
\end{equation}
where $\mathcal{P}_\uparrow^{(i)}=\dyad \uparrow_i$ is the projector onto the higher-lying Rydberg state, and $\mathrm{h.c.}$ denotes the Hermitian conjugate.
To simplify our analysis, couplings to next-nearest neighbors and beyond are neglected throughout this work.

The two outer atoms of the chain couple to the electric field of the corresponding electromechanical oscillator.
This results in a dipolar coupling between the atoms and the oscillators~\cite{Stevenson2016, Gao2011}.
We assume resonant coupling with $\omega=\Delta$. Under this condition, the coupling Hamiltonian is given by
\begin{align}
    H_\mathrm{couple}=J\left(\dyad{\downarrow}{\uparrow}_1 \otimes a^\dagger + \dyad{\downarrow}{\uparrow}_N \otimes b^\dagger + \mathrm{h.c.}\right).
    \label{eq:exc-hamiltonian}
\end{align}
The parameter $J$ quantifies the interaction strength between the mechanical systems and the Rydberg atoms.
For realistic experimental parameters, as discussed in Appendix~\ref{sec:exp-constraints}, the minimal atom-oscillator distance will limit the atom-oscillator coupling $J$ to be smaller than the atom-atom interaction $V$ for typical tweezer distances.
At the same time, the decay rates of the Rydberg atoms $\gamma_\uparrow,\, \gamma_\downarrow$ and oscillators $\kappa$ are orders of magnitude smaller than the coherent interaction strengths $J$ and $V$. Consequently, the coherent dynamics of the system are experimentally accessible on timescales where $\gamma_\downarrow t,\, \gamma_\uparrow t,\, \kappa t \ll 1$.
Furthermore, the decay rates of the Rydberg states can be artificially increased in experiments by laser dressing~\cite{Begoc2025,Schempp2015,Whitlock2019}.
We will analyze the generation of entanglement between the oscillators for both, purely coherent dynamics and dynamics with engineered dissipation in Sec.~\ref{sec:coherent-dynamics} and Sec.~\ref{sec:dissipative-dynamics}, respectively.

To assess the entanglement of the oscillators, we analyze their reduced density matrix $\rho_{ab}$,
which is obtained by performing a partial trace over the atom chain.
In general, the reduced density matrix will represent a mixed state.
To quantify the entanglement of this state, we use the negativity~\cite{Vidal2002,Horodecki2009}, which for our system is calculated as
\begin{equation*}
\label{eq:negativity}
    \mathcal N(\rho_{ab}) = \frac{1}{2}\left(\norm{T_a\tr_\mathrm{chain}(\rho)}_1-1\right).
\end{equation*}
Here, $\rho$ is the density matrix of the full system, $\tr_\mathrm{chain}$ the partial trace over the Rydberg chain, $T_a$ the partial transpose with respect to oscillator $a$ and $\norm{\cdot}_1$ the tracenorm.
The oscillators are entangled if the negativity is larger than zero, with the largest negativity being reached by a maximally entangled state~\cite{Vidal2002}.
One choice for such a maximally entangled state is $\ket{\phi_\mathrm{max}}=\frac{1}{\sqrt{n+1}}\sum_{i=0}^n \ket{i}_a\otimes\ket{i}_b$, with $n$ being an integer number~\cite{Horodecki2009}.
Here, $\ket{i}_{a/b}$ is the Fock state of oscillator $a$ and $b$, respectively. 
This state has negativity $\mathcal{N}(\dyad{\phi_\mathrm{max}}) = n/2$
and the oscillators contain on average $\ev{ a^\dagger a + b^\dagger b}{\phi_\mathrm{max}}=n$ excitations.
In fact, the negativity is bounded from above by
\begin{equation}
    \label{eq:neg-phi-max}
    \mathcal{N} \le \frac{\Big \lceil \ev{a^\dagger a + b^\dagger b} \Big\rceil}{2},
\end{equation}
where \(\lceil\cdot\rceil\) denotes the ceiling function.

This allows us to give an upper bound for the negativity that we can expect in our system. To construct the upper bound, we utilize that the operator
\begin{equation*}
    \mathcal{M} = a^\dagger a + b^\dagger b + \sum_{i=1}^{N} \mathcal{P}_\uparrow^{(i)}
\end{equation*}
commutes with the full Hamiltonian~\eqref{eq:total-hamiltonian} and thus $\mu=\ev{\mathcal M}$ is conserved under coherent evolution. 
This operator counts the number of atoms in the state \up together with the number of excitations in the oscillators.
Throughout this work, we call $\mu$ the total excitation number and refer to atoms in state \up as spin excitations.
It follows that the number of excitations in the initial state of the system defines an upper bound on the maximum reachable negativity during the time evolution with
\begin{equation}
    \label{eq:neg-upper-bound}
    \mathcal N_\text{max} \le \frac{\lceil \mu \rceil}{2}.
\end{equation}
This shows that a higher total number of initial excitations allows for higher negativity values.

\section{Deterministic entanglement generation through coherent evolution}
\label{sec:coherent-dynamics}
In this section, we illuminate the mechanisms that underlie the entanglement generation between the oscillators.
To that end, we focus on the purely coherent dynamics, which are experimentally accessible on timescales $\gamma_\downarrow t,\, \gamma_\uparrow t,\, \kappa t \ll 1$, where dissipation is negligible.
The coherent evolution of the system is governed by the full Hamiltonian~\eqref{eq:total-hamiltonian}.
This Hamiltonian only indirectly describes any effective coupling between the oscillators via the atom-atom and atom-oscillator coupling, see Eqs.~\eqref{eq:chain-hamiltonian},\,\eqref{eq:exc-hamiltonian}.
We now exploit that for typical experimental parameters $J\ll V$.
This allows us to perturbatively derive an effective Hamiltonian that directly couples the oscillators. 
To do so, it is necessary to compute the eigenstates of $H_\mathrm{chain}$.
For arbitrary chain lengths $N$, this is possible using the Jordan-Wigner transformation~\cite{Coleman2008}. 
To illustrate the principle, we consider a chain of two atoms, where the eigenstates are the spin-singlet state $\ket S=(\ket{\uparrow\downarrow}-\ket{\downarrow\uparrow})/\sqrt 2$ and the spin-triplet states $\ket{\uparrow\uparrow}$, $\ket{\downarrow\downarrow}$ and $\ket T =(\ket{\uparrow\downarrow}+\ket{\downarrow\uparrow})/\sqrt 2$.
Expressing the chain Hamiltonian~\eqref{eq:chain-hamiltonian} in its eigenbasis and employing a Schrieffer-Wolff transformation~\cite{Schrieffer1966,Landi2024}, we obtain the effective Hamiltonian
\begin{align}
    H_\mathrm{eff}=
    &- \frac{J^2}{V} \left(a^\dagger b + ab^\dagger\right)\left(\dyad{\downarrow\downarrow}+\dyad{\uparrow\uparrow}-\dyad S -\dyad T\right)\nonumber\\
    &-\frac{J^2}{V}\left[\left({a^\dagger}^2+{b^\dagger}^2\right) \dyad{\downarrow\downarrow}{\uparrow\uparrow} +\left(a^2+b^2\right) \dyad{\uparrow\uparrow}{\downarrow\downarrow} \right]\nonumber\\
    &+\left(V+\frac{J^2 \mathcal M}{V}\right)(\dyad T-\dyad S)+\mathcal M \omega.
    \label{eq:eff-hamiltonian}
\end{align}
Here, the third line represents the energy of the basis states of the system, while the first two lines contain qualitatively different coupling terms.
The first line describes a direct oscillator–oscillator coupling that generates entanglement, where the sign of the coupling depends on the state of the chain.
The second line describes an indirect coupling at higher order, which involves exchange of excitation pairs between oscillators and chain.

To illustrate the roles of these coupling terms, we consider two different initial states.
The first state is $\ket{\psi_1} = \ket{1}_a\otimes\ket{S}\otimes\ket{0}_b$, where one excitation resides in oscillator $a$ and the chain is in the singlet state.
The singlet state contains one spin excitation.
The second state we consider is $\ket{\psi_2} = \ket{0}_a\otimes\ket{\uparrow\uparrow}\otimes\ket{0}_b$, where both spins are initially excited, and the oscillators are in the ground state $\ket 0$.
Both states have the same total number of excitations $\mu=2$ and consequently have the same upper bound of the maximum reachable negativity according to Eq.~\eqref{eq:neg-upper-bound}.

For these initial states, the time evolution of the oscillator excitation number as well as the negativity under the effective Hamiltonian~\eqref{eq:eff-hamiltonian} can be calculated analytically. They are given by
\begin{align*}
    \ev{a^\dagger a}{\psi_1}=& 1-\ev{b^\dagger b}{\psi_1} = \frac{1}{2}\left[1+\cos \left(\frac{\tau}{2\sqrt{2}}\right)\right],\\
    \ev{a^\dagger a}{\psi_2}=&\ev{b^\dagger b}{\psi_2} = 1-\cos^4\left(\frac{\tau}{2}\right),\\
    \mathcal N (\dyad{\psi_1})=&\frac{1}{2}\abs{\sin\left(\frac{\tau}{2\sqrt{2}}\right)},\\
    \mathcal N (\dyad{\psi_2})=&\frac{1}{2}\big [1-\cos(\tau)\big ]\left|\sin\left(\tau\right)\right|\\
    &+\frac{1}{8}\big [1+\cos(\tau)\big ]\\
    &\times\left(\sqrt{5\cos^2(\tau)-6\cos(\tau)+5}-1-\cos(\tau)\right),
\end{align*}
where $\tau=\frac{2\sqrt{2}J^2}{V}t$.
In Fig.~\ref{fig:cartoon}(b,~c), we present the analytic results together with the numerical solution of the full Hamiltonian~\eqref{eq:total-hamiltonian}. The effective model captures the overall dynamics.
For both initial states, entanglement between the oscillators gets periodically created and destroyed during the evolution of the system. 
We find in Fig.~\ref{fig:cartoon}(b,~c) that a higher negativity is reached during the time evolution of $\ket{\psi_2}$ compared to $\ket{\psi_1}$.
This can be understood by considering the allowed exchange processes for the two initial states.
For $\ket{\psi_1}$, only single-excitation exchange between the oscillators via the first term in $H_\mathrm{eff}$ is possible, such that the number of excitations $\ev{a^\dagger a + b^\dagger b}$ in the oscillators remains fixed at one.
Hence, a stricter bound on the negativity, $\mathcal N(\dyad{\psi_1})\le1/2$, is imposed according to Eq.~\eqref{eq:neg-phi-max}.
On the contrary, for $\ket{\psi_2}$, the exchange of excitation pairs with the chain via the second term in $H_\mathrm{eff}$ allows the system to transfer the two excitations, initially in the atom chain, into the oscillators. Although this pair exchange still couples the oscillators only indirectly via the chain, it enables the system to reach entangled states with more excitations in the oscillators.
This leads to the larger negativity reached for this initial state.

\section{Probabilistic entanglement generation through dissipative processes}
\label{sec:dissipative-dynamics}

For the creation of correlated oscillator states, one would ideally stop the coherent dynamics when the maximum negativity is reached. While this is in principle possible, e.g., by removing the atoms with repulsive optical tweezers, this requires exact timing and precise knowledge of the system parameters. 
In the following, we discuss a scheme where artificially enhanced dissipation in the atomic chain is exploited to generate an entangled oscillator state and, at the same time, to stop the coherent dynamics of entanglement creation and destruction automatically.

To illustrate this idea, we consider a separable initial state of the form
\begin{align}
    \ket{\psi_\mathrm{init}}= \ket{0}_a\otimes\ket{\uparrow}^{\otimes N} \otimes \ket{0}_b,
    \label{eq:initial_state_dissipative}
\end{align}
where all atoms are in the Rydberg state \up and the oscillators are in the vacuum state.
Over the course of the dissipative time-evolution --- which we discuss in detail below --- this state will eventually evolve into a product state, where the atomic chain is found in $\ket{g}^{\otimes N}$. We refer to this state as the final state. 
From this point onward, the coupling between the atom chain and the oscillators stops and the latter merely continue the evolution under their individual harmonic oscillator Hamiltonians, which can neither create nor remove correlations. The precise sequence and timing of the atomic decay events --- which form a so-called quantum jump trajectory --- are, however, probabilistic. 
This also means that the entanglement of the final oscillator state, which is reached when all atoms have decayed, is probabilistic.

\subsection{Mechanism of dissipative entanglement generation}
\label{sec:single-trajectory-dynamics}
\begin{figure}[t]
    \centering
    \includegraphics[width=\linewidth]{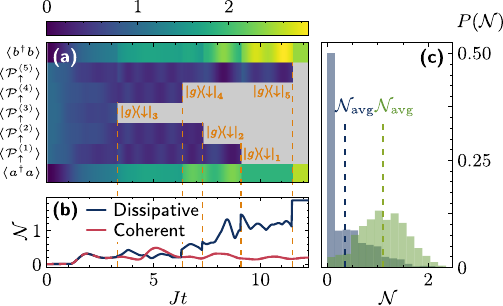}
    \caption{\textbf{Entanglement generation through atom chain dissipation.}
    \textbf{(a):}~Time evolution of excitations in the sites of the spin chain $\ev*{\mathcal{P}_\uparrow^{(i)}}$ and the number of excitations in the oscillators $\ev*{a^\dagger a},\,\ev*{b^\dagger b}$ for a selected single trajectory and system parameter \(V = 3J\), \(\gamma_\downarrow = \gamma_\uparrow = 0.2J\) and $\kappa=0$. The system is initially prepared in the state \(\ket{\psi_\mathrm{init}} = \ket{0}_a \otimes \ket{\uparrow\uparrow\uparrow\uparrow\uparrow} \otimes \ket{0}_b\).
    Atom decay is marked with the corresponding jump operator, with $\sqrt{\gamma_\downarrow}$ omitted. Atoms that have decayed to the ground state are colored gray. 
    \textbf{(b):}~Time evolution of the oscillator negativity \(\mathcal{N}\) for the trajectory shown in (a) (blue) in comparison to the purely coherent evolution (red) for the same parameters. 
    \textbf{(c):}~Probability distributions $P(\mathcal N)$ of final-state negativities $\mathcal{N}$ for $2000$ trajectories using the parameters from (a) with $\gamma_\uparrow=0.2J$ (blue) and $\gamma_\uparrow=0$ (green). The averages of the negativity distributions $\mathcal N_\mathrm{avg}$ are depicted by vertical dashed lines.}
    \label{fig:cartoon-entanglement}
\end{figure}
We simulate quantum trajectories of the full system using the quantum jump method discussed in Refs.~\cite{Dalibard1992,Plenio1998}. 
In this framework, the evolution of the initial state, defined in Eq.~(\ref{eq:initial_state_dissipative}), is governed by the non-Hermitian Hamiltonian $\mathcal H = H-\tfrac{i}{2}\sum_{\alpha} L_\alpha^\dagger L_\alpha$, with the jump operators
\begin{align*}
    L_\alpha\in\Big\{
        &\sqrt{\gamma_\uparrow}\dyad{g}{\uparrow}_1,\;...,\;\sqrt{\gamma_\uparrow}\dyad{g}{\uparrow}_N,\nonumber\\
        &\sqrt{\gamma_\downarrow}\dyad{g}{\downarrow}_1,\;...,\;\sqrt{\gamma_\downarrow}\dyad{g}{\downarrow}_N,\;
        \sqrt{\kappa}a,\;\sqrt{\kappa}b\Big\}.
\end{align*}
Here, the first $2N$ terms capture the decay from the Rydberg states \up and \down to the state $\ket{g}$, while the last two describe the decay of the oscillators. These jumps occur at random times, reflecting the probabilistic nature of the quantum trajectories.

To elucidate the role of the Rydberg atom decay in the evolution of the system, we consider a system with $N=5$ atoms, with an interaction strength $V=2J$ and decay rates $\gamma_\uparrow=\gamma_\downarrow=0.2J$ for the Rydberg state.
For simplicity, we switch off the oscillator decay ($\kappa=0$).
In Fig.~\ref{fig:cartoon-entanglement}(a), we depict the time evolution of the number of excitations in the oscillators and the number of spin excitations within the atomic chain for a single trajectory.
We compare the time evolution of the negativity of the reduced oscillator state for this selected trajectory to that of the fully coherent evolution without decay of the same system in Fig.~\ref{fig:cartoon-entanglement}(b).
A striking observation is that the negativity during this dissipative evolution reaches a significantly higher value compared to the purely coherent evolution until $Jt=12$.
This can be explained as follows: In the selected trajectory, the first and all following decays take place via the decay channel $\ket{\downarrow}\rightarrow\ket{g}$.
The corresponding collapse operators $\sqrt{\gamma_\downarrow}\dyad{g}{\downarrow}_{k}$, with $k$ the index of the atom, do not change the total number of excitations $\mu$ in the system.
Hence, during the evolution, all initial spin excitations are converted to oscillator excitations, as shown in Fig.~\ref{fig:cartoon-entanglement}(a).
Thus, the negativity of the final state reached by this trajectory can approach the maximum negativity bound in accordance with Eq.~\eqref{eq:neg-phi-max}.
In the absence of decay, excitations are continuously exchanged between chain and oscillators, with some excitations present in the chain at all times. Hence, the negativity of this coherent evolution remains lower.

In the selected example trajectory, only decay from state \down occurred. 
In contrast, any decay from state \up reduces the total number of excitations $\mu$ by one.
Hence, the achievable negativity also reduces according to Eq.~\eqref{eq:neg-upper-bound}.
In particular, if all atoms decay via this decay channel, the total number of excitations contained within the system is $\mu=0$, resulting in a completely unentangled final state.
These considerations suggest that, to maximize the negativity reached by the system, it is beneficial to increase $\gamma_\downarrow$, while keeping $\gamma_\uparrow$ unchanged at its minimal value given by spontaneous decay of the Rydberg state.

To further investigate the impact of decay from state \up on the entanglement generation, we now analyze a collection of trajectories.
Given that each trajectory is formed through a probabilistic sequence of decay events, the negativity $\mathcal{N}$ of the final state --- which is reached as soon as all atoms are in the state $\ket{g}$ --- is a random variable. 
It is characterized by a distribution $P(\mathcal N)$, whose shape can be controlled. 
This is shown in Fig.~\ref{fig:cartoon-entanglement}(c), where we plot the distributions for two different rates of decay from the state \up: $\gamma_\uparrow=0.2J$ and $\gamma_\uparrow=0$.
The distribution for $\gamma_\uparrow=0.2J$ features a long tail extending to high negativity values, which are close to the upper negativity bound, $\mathcal N\le2.5$.
However, about half of the $2000$ sampled trajectories yield near zero negativity. 
When $\gamma_\uparrow=0$, the distribution has fewer trajectories with near-zero negativity. In addition, higher negativity values are reached more often.
Consequently, the average final-state negativity $\mathcal N_\mathrm{avg}$ is higher when $\gamma_\uparrow=0$, as marked in Fig.~\ref{fig:cartoon-entanglement}(c).

\subsection{Parameter dependence of final-state entanglement}
\label{sec:decay-rate-engineering}
\begin{figure}[t]
    \centering
    \includegraphics[width=\linewidth]{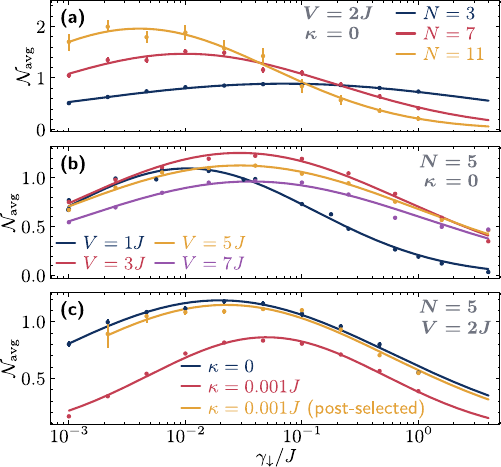}
    \caption{\textbf{Parameter dependence of average final-state negativity.} 
    \textbf{(a):}~The average final-state negativity $\mathcal N_\mathrm{avg}$ as a function of $\gamma_\downarrow$ for different atom numbers $N$ and fixed parameters $V = 2J$, $\kappa = 0$, and $\gamma_\uparrow = 0.001J$. Averages are taken over 1000 trajectories for $N = 3$ and $N = 7$, and over 100 trajectories for $N = 11$.
    \textbf{(b):}~$\mathcal N_\mathrm{avg}$ as a function of $\gamma_\downarrow$ for different coupling strengths $V$ with $N = 5$, $\kappa = 0$, and $\gamma_\uparrow = 0.001J$. Each data point is averaged over $1000$ trajectories.
    \textbf{(c):}~Comparison of two cases, with ($\kappa=0.001J$) and without oscillator decay. Parameters are $V = 2J$, $N = 5$, and $\gamma_\uparrow = 0.001J$. 
    For the case with oscillator decay, we also show results after post-selecting trajectories that complete within a cutoff time of $0.3/\kappa$.
    For $\gamma_\downarrow = 0.001J$ the data point is missing because no trajectory satisfied the post-selection criterion. Averages are taken over $10^4$ trajectories.
    In all panels, log-normal curves are fitted to the data.
    }
    \label{fig:alpha-and-n-and-ge-scan}
\end{figure}
Having shown that $\gamma_\uparrow$ strongly influences the generated entanglement, we next explore how the parameters $N$, $V$, $\gamma_\downarrow$, and $\kappa$ impact the negativity reached by the system. Here, we consider experimentally feasible parameter ranges, as discussed in Appendix~\ref{sec:exp-constraints}.
For this investigation, we set $\gamma_\uparrow=0.001J$, which, for typical experimental parameters, corresponds to the natural decay rate of Rydberg states.
We focus in the following specifically on the average final-state negativity $\mathcal N_\mathrm{avg}$, which is calculated from histograms of the type shown in Fig.~\ref{fig:cartoon-entanglement}(c). 
In Fig.~\ref{fig:alpha-and-n-and-ge-scan}, we show the average negativity $\mathcal{N}_\mathrm{avg}$ as a function of $\gamma_\downarrow$. 
We find that for all considered parameters, the resulting scaling with $\gamma_\downarrow$ is well described by the Gaussian function on a logarithmic scale $\mathcal{N}_\mathrm{avg}(\gamma_\downarrow)=\tfrac{A}{\gamma_\downarrow} \exp(-\frac{(\ln \gamma_\downarrow -\nu)^2}{2\sigma^2})$, where $A$, $\nu$ and $\sigma$ are fit parameters.
This shape reflects the interplay of two competing mechanisms:
On the one hand, increasing $\gamma_\downarrow$ reduces the ratio of decay from the state \up compared to \down.
Thereby, unwanted decays from state \up, which would remove excitations from the system, become less likely and the average negativity increases.
On the other hand, a larger decay rate $\gamma_\downarrow$ results in an earlier first decay event, which splits the chain and divides the full system into two uncoupled subsystems.
After such an event, each sub-chain is only coupled to one of the mechanical oscillators. Quantum correlations can still build up within each of these two subsystems, but not anymore between them. Therefore, the oscillators cannot end up in an entangled state if the two subsystems are not already entangled with each other prior to the first decay. Consequently, a too high $\gamma_\downarrow$ suppresses the entanglement generation. 
In particular, the optimal decay rate $\gamma_\downarrow$ --- at which the maximum of $\mathcal{N}_\mathrm{avg}$ is reached --- is smaller than $J$, such that the system has enough time to transfer excitations into the oscillators and build up correlations.
This highlights the importance of engineering $\gamma_\downarrow$ for the dissipative entanglement generation.

In Fig 3(a) we investigate the role of the number of atoms $N$ in the Rydberg atom chain. We find that increasing $N$ results in larger maximal $\mathcal{N}_{\mathrm{avg}}$.
This is because the chosen initial state, defined in Eq.~\eqref{eq:initial_state_dissipative}, contains more excitations for longer chains, which increases the reachable negativity according to Eq.~\eqref{eq:neg-upper-bound}.
However, a larger $N$ also increases the number of possible decay channels, resulting in an earlier first decay event. 
Additionally, the time for initial correlation buildup scales with $N$, as excitations must travel through a longer chain for correlations to form between the two oscillators.

Thus, the optimal decay rate $\gamma_\downarrow$ shifts towards lower values for higher $N$.
This trend suggests that increasing the number of atoms $N$, while keeping the atomic decay rates, $\gamma_\uparrow$ and $\gamma_\downarrow$, minimal, may be best. We stress that this is only correct for the idealized case of infinitely long-lived mechanical oscillators ($\kappa=0$), for which the available time for correlation buildup is limited only by the Rydberg decay rates.

Next, we study the role of the interaction strength $V$, while keeping $N=5$ and $\kappa=0$ fixed.
Among the investigated values of $V$, the highest average final-state negativity $\mathcal{N}_{\mathrm{avg}}$ is reached for $V=3J$ and $\gamma_\downarrow\approx 0.03J$, as shown in Fig.~\ref{fig:alpha-and-n-and-ge-scan}(b). 
For $V=1J$, the maximum $\mathcal{N}_\mathrm{avg}$ shifts towards lower $\gamma_\downarrow$ and is overall smaller.
The reason is that a smaller $V$ decelerates the excitation hopping in the chain. This results in a slower correlation buildup that is more likely to be interrupted by the first decay.
Interestingly, we also observe a reduction of the maximum of $\mathcal{N}_{\mathrm{avg}}$ for larger $V=5J,\,7J$.
This we understand by considering the effective Hamiltonian~\eqref{eq:eff-hamiltonian}, which we derived for $N=2$ atoms and $V\gg J$. We see that the chain mediates a coherent coupling among the oscillator with effective rate $J^2/V$.
Thus, the process of transporting spin excitations out of the chain into the oscillators takes longer the larger $V$ becomes.
As a consequence, the initially fully excited spin chain remains longer close to fully excited. 
This, in turn, increases the probability of the detrimental decay from state \up in the spin chain, resulting in a reduced average negativity.

Up to this point, we have considered idealized oscillators without any internal decay ($\kappa=0$). To illustrate the impact of a finite oscillator lifetime, we compare in Fig.~\ref{fig:alpha-and-n-and-ge-scan}(c) the ideal system with one where $\kappa=0.001J$; a value which is achievable with current state-of-the-art devices, see Appendix~\ref{sec:exp-constraints}.
We find that excitation loss due to oscillator decay yields a reduction of $\mathcal{N}_\mathrm{avg}$ in accordance to Eq.~\eqref{eq:neg-upper-bound}.
The reduction becomes more severe for smaller $\gamma_\downarrow$, which also results in a shift of the optimal $\gamma_\downarrow$ to higher values.
This is because the time it takes for all atoms to decay increases as the value of $\gamma_\downarrow$ decreases.
Evidently, the longer this time is, the more likely it is for the oscillators to decay to an uncorrelated state.
To remedy this, one can post-select only those trajectories where all atoms decay to state $\ket{g}$ in a time that is short compared to the oscillator lifetime.
Experimentally, the occupation of state \g can be measured sufficiently fast compared to the oscillator lifetime to implement such a post selection strategy~\cite{Falconi2025}.
In Fig.~\ref{fig:alpha-and-n-and-ge-scan}(c), we present a post-selected dataset, to which only trajectories contribute in which all atoms decay to the state $\ket{g}$ in under $0.3/\kappa$.
This post-selection increases $\mathcal{N}_\mathrm{avg}$ to nearly the same value as in the case without oscillator decay, at the cost of the fraction of accepted trajectories reducing drastically when $\gamma_\downarrow$ is small. For example, for $\gamma_\downarrow=0.001J$, which corresponds to the natural decay rate, no trajectory out of $10,000$ satisfies the post-selection requirement.
In contrast, the acceptance fraction for the optimal decay rate $\gamma_\downarrow=0.02J$ is $6.3\%$.

\section{Conclusion and outlook}
In this work, we demonstrated how a chain of Rydberg atoms can mediate entanglement between two micro-electromechanical oscillators. In particular, we shed light on the influence of both coherent and dissipative effects during this process. While dissipation renders entanglement generation probabilistic,
we showed that under certain circumstances, decay events may even enhance the creation of quantum correlations.

An interesting topic for future investigation concerns exploiting the knowledge of the atomic decay events. For example, if it was possible to experimentally identify the decay time and decay channel of the individual atoms, one could --- in the absence of oscillator decay --- construct the conditioned pure state of the oscillators \cite{Gammelmark2013}. This is a somewhat extreme and practically challenging scenario. However, as shown towards the end of the previous section, even simpler post-selection protocols yield larger average negativities. 
Searching for more efficient post-selection schemes thus seems like a worthwhile endeavor.

A further pathway for boosting oscillator entanglement is to introduce time-dependent coherent coupling protocols, as in Rydberg atom tweezer quantum simulators and computers~\cite{Evered2023,Scholl2022}. In these studies, it would be interesting to lift the approximation of the nearest-neighbor interaction among the atoms and to incorporate power-law potentials. This may improve the robustness of the entanglement generation, since the loss of an atom would not automatically lead to two decoupled sub-chains. In a similar vein, using two-dimensional Rydberg tweezer arrays to mediate interactions between the oscillators could enhance robustness to atom loss.

\section*{Acknowledgments}
The research leading to these results has received funding from the Deutsche Forschungsgemeinschaft (DFG, German Research Foundation) under Project No. 449905436, Research Unit FOR 5413/1, Grant No. 465199066, and Germany’s Excellence Strategy --- Cluster of Excellence Matter and Light for Quantum Computing (ML4Q) EXC 2004/1 --- 390534769, from the European Union’s Horizon 2020 program under the ERC grants SUPERWAVE (Grant No.\ 101071882) and OPEN-2QS (Grant No.\ 101164443). This work was also supported by the QuantERA II programme (project CoQuaDis, DFG Grant No. 532763411) that has received funding from the EU H2020 research and innovation programme under GA No. 101017733.

\appendix
\section{Experimental constraints on parameter choices}
\label{sec:exp-constraints}
The parameter ranges explored in the main text are informed by experimental consideration that we outline here.
Among the various micro-electromechanical oscillator platforms capable of achieving high frequencies and long lifetimes~\cite{OConnell2010,Chu2017,Han2016,Chu2020,Luo2025}, we take guidance from high-overtone bulk acoustic resonators (HBARs). These devices are particularly promising for quantum acoustic experiments due to their exceptionally high quality factors. State-of-the-art HBARs typically operate at resonance frequencies $\omega$ around \SI{5}{\giga\hertz}, with decay rates around \SI{1}{\kilo\hertz}~\cite{Gokhale2020,Yang2024,vonLupke2024}.

This resonance frequency can be matched to transitions between Rydberg states with principal quantum numbers around $90$, which exhibit natural lifetimes on the order of \SI{1}{\milli\second}~\cite{Sibalic2017}. Off-resonant dressing with lower-lying states can strongly enhance the decay rate. For instance, in rubidium, dressing the Rydberg state with the first excited state allows the increase of the decay rate of up to \(\sim\)\SI{3}{\mega\hertz}.

The large dipole matrix elements between such high-lying Rydberg states enable strong coupling to the evanescent field of the HBAR. For realistic atom-surface separations of \SIrange{20}{30}{\micro\meter}, we estimate the coupling strength $J$ to be in the range of \SIrange{0.5}{1.2}{\mega\hertz} using the calculation methods and material constants outlined in~\cite{Gao2011, vonlupke2023}. Similarly, the dipole-dipole interaction $V$ between Rydberg atoms is substantial, with \(C_3\) coefficients on the order of $\sim$\SI{80}{\giga\hertz\micro\meter^3}. This yields interatomic couplings of roughly \SIrange{1}{10}{\mega\hertz} for separations of \SIrange{20}{40}{\micro\meter}~\cite{Sibalic2017,Barredo2015,deLeseleuc2019}.
For simplicity, we fix the atom-oscillator coupling strength to be \(J = \SI{1}{\mega\hertz}\) and scale the other parameters accordingly for the parameter scans presented in Sec~\ref{sec:decay-rate-engineering}.

\section{Zero-temperature environment}
\label{sec:zero-temperature}
We assume a zero temperature environment for the treatment of dissipation in the system.
As a consequence, heating of the oscillators and blackbody-induced transitions between Rydberg states vanish.

In experiments, temperatures down to $T\approx\SI{10}{\milli\kelvin}$ are typically reached within a $^3\text{He}/^4\text{He}$ dilution refrigerator~\cite{Pobell2007}.
At these temperatures, thermal heating of an oscillator mode at the considered frequencies around $\SI{5}{\giga\hertz}$ is negligible, since $k_\mathrm BT\ll\hbar\omega$, where $k_\mathrm B$ is the Boltzmann constant. Furthermore, blackbody-induced transitions are also strongly suppressed and are negligible compared to spontaneous decay~\cite{Sibalic2017}.

\bibliography{biblio}
\end{document}